\documentclass[11pt]{article}%
\usepackage{amsfonts}
\usepackage{amsmath}
\usepackage{amssymb}
\usepackage{graphicx}%
\begin{document}

\title{Consequences from conservation of the total density of the universe during the expansion}
\author{Dimitar Valev\\\textit{Stara Zagora Department, Space and Solar-Terrestrial}
\textit{Research}\\\textit{Institute,} \textit{Bulgarian Academy of Science, 6000 Stara Zagora,
Bulgaria}}
\maketitle

\begin{abstract}
The recent Cosmic Microwave Background (\textit{CMB}) experiments have shown
that the average density of the universe is close to the critical one and the
universe is asymptotically flat (Euclidean). Taking into account that the
universe remains flat and the total density of the universe $%
\Omega
_{0}$ is conserved equal to a unit during the cosmological expansion, the
Schwarzschild radius of the observable universe has been determined equal to
the Hubble distance $R_{s}=2GM/c^{2}=R\sim c/H$, where $M$ is the mass of the
observable universe, $R$ is the Hubble distance and $H$ is the Hubble
constant. Besides, it has been shown that the speed of the light $c$ appears
the parabolic velocity for the observable universe $c=\sqrt{2GM/R}=v_{p}$ and
the recessional velocity $v_{r}=Hr$ of an arbitrary galaxy at a distance
$r>100$ $Mps$ from the observer, is equal to the parabolic velocity for the
sphere, having radius $r$ and a centre, coinciding with the observer. The
requirement for conservation of $\Omega_{0}=1$ during the expansion enables to
derive the Hoyle-Carvalho formula for the mass of the observable universe
$M=c^{3}/(2GH)$ by a new approach.


\end{abstract}

\section{Introduction}

The problem for the average density of the universe $\overline{\rho}$ acquires
significance when it was shown that the General Relativity allows to reveal
the geometry and evolution of the universe by simple cosmological models
\cite{Friedmann, Lemaitre, Einstein}. Crucial for the geometry of the universe
appears the dimensionless total density of the universe $\Omega_{0}%
=\overline{\rho}/\rho_{c}$, where $\overline{\rho}$ is the average density of
the universe and $\rho_{c}$ is the critical density of the universe. In the
case of $%
\Omega
_{0}<1$ (open universe) the global spatial curvature is negative and the
geometry of the universe is hyperbolic and in the case of $%
\Omega
_{0}>1$ (closed universe) the curvature is positive and the geometry is
spherical. In the special case of $%
\Omega
_{0}=1$ (flat universe) the curvature is zero and the geometry is Euclidean.
Until recently scarce information has been available about the density and
geometry of the universe. The most reliable determination of the total density
$\Omega_{0}$ is by measurements of the dependence of the anisotropy of the
Cosmic Microwave Background ($CMB$) upon the angular scale. The recent results
have shown that $\Omega_{0}\approx1\pm\Delta\Omega_{0}$, where the error
$\Delta\Omega_{0}$ decreases from 0.10 \cite{de Bernardis, Balbi} to 0.02
\cite{Spergel}, i.e. the density of the universe is close to the critical one
and the universe is asymptotically flat (Euclidean).

The fact that $\Omega_{0}$ is so close to a unit is not accidental since only
at $%
\Omega
_{0}=1$ the geometry of the universe is flat and the flat universe was
predicted from the inflationary theory \cite{Guth}. The total density
$\Omega_{0}$ includes density of baryon matter $\Omega_{b}\approx0.05$, cold
dark matter $\Omega_{c}\approx0.22$ \cite{Peacock} and dark energy
$\Omega_{\Lambda}\approx0.73$ \cite{Hinshaw}, producing an accelerating
expansion of the universe \cite{Riess, Perlmutter}. The found negligible $CMB$
anisotropy $\delta T/T\sim10^{-5}$ indicates that the early universe was very
homogeneous and isotropic \cite{Bennett}. Three-dimensional maps of the
distribution of galaxies corroborate homogeneous and isotropic universe on
large scales greater than $100$\ $\ Mps$ \cite{Shectman, Stoughton}.

\section{Consequences from conservation of the total density of the universe
during the expansion}

The flat geometry of the universe allows to solve some cosmological problems
in the Euclidean space. The finite time of the cosmological expansion $H^{-1}$
(age of the universe) and the finite speed of the light $c$ set a finite
particle horizon $R\sim cH^{-1}$ beyond which no material signals reach the
observer. Therefore, for an observer in an arbitrary location, the universe
appears a three-dimensional, homogeneous and isotropic sphere having finite
\textquotedblleft radius\textquotedblright\ (particle horizon) equal to the
Hubble distance $R\sim cH^{-1}$, where $H\approx70$ $km$ $s^{-1}$ $Mps^{-1}$
\cite{Mould} is the Hubble constant and $H^{-1}\approx1.37\times10^{10}$
$years$ is the Hubble time (age of the universe).

The fact that the total density of the universe $\Omega_{0}$ is close to a
unit is fundamental since only $\Omega_{0}=\overline{\rho}/\rho_{c}=1$
supplies flat geometry of the universe. There are no arguments to assume the
recent epoch privileged in relation to the other epochs; therefore, the
universe always remains flat, and \textit{the total density of the universe }$%
\Omega
_{0}$\textit{\ is conserved equal to a unit during the cosmological expansion}:%

\begin{equation}
\Omega_{0}=\frac{\overline{\rho}}{\rho_{c}}=1 \label{Eqn1}%
\end{equation}

The critical density of the universe \cite{Peebles} is determined from
equation (\ref{Eqn2}):%

\begin{equation}
\rho_{c}=\frac{3H^{2}}{8\pi G}\approx9.5\times10^{-27}%
\operatorname{kg}%
\operatorname{m}%
^{-3} \label{Eqn2}%
\end{equation}

where $G$ is the universal gravitational constant.

Considering $\overline{\rho}=3M/(4\pi R^{3})$, where $M$ and $R$ are the mass
and the Hubble distance (\textquotedblleft radius\textquotedblright) of the
observable universe, and replacing $\rho_{c}$ with expression (\ref{Eqn2}) in
(\ref{Eqn1}) we obtain:%

\begin{equation}
\frac{2MG}{R^{3}H^{2}}=1 \label{Eqn3}%
\end{equation}

Replacing $H\sim cR^{-1}$ in (\ref{Eqn3}) we obtain:%

\begin{equation}
R=\frac{2GM}{c^{2}} \label{Eqn4}%
\end{equation}

Obviously, (\ref{Eqn4}) appears the formula for the Schwarzschild radius
\cite{Schwarzschild} of the mass of the observable universe $M$. Therefore,
\textit{the Schwarzschild radius of the observable universe }$R_{s}%
$\textit{\ is equal to the Hubble distance }$R_{s}=R\sim cH^{-1}\sim
1.37\times10^{10}$\textit{\ light years.}

From (\ref{Eqn4}) we find:%

\begin{equation}
c=\sqrt{\frac{2GM}{R}} \label{Eqn5}%
\end{equation}

Evidently, (\ref{Eqn5}) is the formula of the parabolic velocity for the
Hubble sphere, i.e. the sphere having mass $M$ and a radius, equal to the
Hubble distance $R\sim cH^{-1}$. Therefore, \textit{the speed of the light
}$c$\textit{\ appears the parabolic velocity }$v_{p}$\textit{\ for the
observable universe}.

Below, we find that the recessional velocity $v_{r}=Hr$ of an arbitrary galaxy
at a distance $r>100$ $Mps$ from the observer is equal to the parabolic
velocity for a sphere, having radius $r$ and a centre, coinciding with the
observer. As mentioned at the end of the Introduction, the universe is
homogeneous and isotropic on large scales greater than $100$ $Mps$. Therefore,
the average density $\rho_{r}$ of a sphere having radius $r>100$ $Mps$ is
equal to the average density of the universe $\overline{\rho}$:%

\begin{equation}
\rho_{r}=\frac{3m}{4\pi r^{3}}=\overline{\rho}\approx\rho_{c}=\frac{3H^{2}%
}{8\pi G} \label{Eqn6}%
\end{equation}

where $m$ is the mass of the total matter in the sphere.

We find from equation (\ref{Eqn6}):%

\begin{equation}
H=\sqrt{\frac{2Gm}{r^{3}}} \label{Eqn7}%
\end{equation}

Replacing $H$ in the Hubble law $v_{r}=Hr$ we obtain the recessional velocity
of a galaxy:%

\begin{equation}
v_{r}=Hr=\sqrt{\frac{2Gm}{r}} \label{Eqn8}%
\end{equation}

Equation (\ref{Eqn8}) coincides with the formula for the parabolic velocity of
a sphere, having radius $r$ and a centre, coinciding with the observer.

Finally, the requirement for conservation of the total density of the universe
equal to a unit during the expansion allows to estimate the total mass of the
observable universe $M$. Actually, replacing $R\sim cH^{-1}$ in (\ref{Eqn3})
we find:%

\begin{equation}
M=\frac{c^{3}}{2GH}\approx8.8\times10^{52}%
\operatorname{kg}
\label{Eqn9}%
\end{equation}

Obviously, this mass is close to the mass of the Hubble sphere $M_{H}$:%

\begin{equation}
M_{H}=\frac{4}{3}\pi R^{3}\overline{\rho}\sim\frac{4\pi c^{3}\rho_{c}}{3H^{3}%
}=\frac{c^{3}}{2GH} \label{Eqn10}%
\end{equation}

Formula (\ref{Eqn9}) has been derived independently by dimensional analysis
without consideration of the average density of the universe in \cite{Valev
2009, Valev 2010} and practically coincides with the Hoyle-Carvalho formula
for the mass of the universe \cite{Hoyle, Carvalho}, obtained by a totally
different approach.

\section{Conclusions}

The recent \textit{CMB} experiments have shown that the average density of the
universe is close to the critical one and the universe is asymptotically flat.
The flat geometry of the universe allows to solve some cosmological problems
in the Euclidean space. Taking into account that the universe remains flat and
the total density of the universe $%
\Omega
_{0}$ is conserved equal to a unit during the expansion, the Schwarzschild
radius of the observable universe has been determined equal to the Hubble
distance $R_{s}=2GM/c^{2}=R\sim cH^{-1}$, and the speed of the light $c$
appears the parabolic velocity for the observable universe $c=\sqrt
{2GM/R}=v_{p}$. Besides, the recessional velocity $v_{r}=Hr$ of an arbitrary
galaxy at a distance $r>100$ $Mps$ from the observer, is equal to the
parabolic velocity of a sphere, having radius $r$ and a centre, coinciding
with the observer.

The requirement for conservation of $\Omega_{0}=1$ during the cosmological
expansion enables to derive the Hoyle-Carvalho formula for the mass of the
observable universe $M=c^{3}/(2GH)$ by a new approach.

\end{document}